\documentclass[preprint, superscriptaddress, showpacs,preprintnumbers,amsmath,amssymb,pra]{revtex4}
\usepackage{graphicx}

\begin{document}

\thispagestyle{empty}

\title{Low-temperature behavior of the Casimir free energy and entropy
of metallic films}

\author{
G.~L.~Klimchitskaya}
\affiliation{Central Astronomical Observatory at Pulkovo of the
Russian Academy of Sciences,
Saint Petersburg,
196140, Russia}
\affiliation{Institute of Physics, Nanotechnology and
Telecommunications, Peter the Great Saint Petersburg
Polytechnic University, Saint Petersburg, 195251, Russia}

\author{
V.~M.~Mostepanenko}
\affiliation{Central Astronomical Observatory at Pulkovo of the Russian
Academy of Sciences, Saint Petersburg,
196140, Russia}
\affiliation{Institute of Physics, Nanotechnology and
Telecommunications, Peter the Great Saint Petersburg
Polytechnic University, Saint Petersburg, 195251, Russia}
\affiliation{Kazan Federal University, Kazan, 420008, Russia}

\begin{abstract}
We derive an analytic behavior of the Casimir free energy, entropy and
pressure of metallic films in vacuum at low temperature. It is shown
that this behavior differs significantly depending on whether the plasma
or the Drude model is used to describe the dielectric properties of film
metal. For metallic films described by the lossless plasma model the
thermal corrections to the Casimir energy and pressure drop to zero
exponentially fast with increasing film thickness. There is no classical
limit in this case. The Casimir entropy satisfies the Nernst heat theorem.
For metallic films with perfect crystal lattices described by the Drude
model the Casimir entropy at zero temperature takes a nonzero value
depending on the parameters of a film, i.e., the Nernst heat theorem is
violated. The Casimir entropy at zero temperature is positive, as opposed
to the case of two metallic plates separated with a vacuum gap, where it
is negative if the Drude model is used. Possible applications of the
obtained results in investigations of stability of thin films are
discussed.
\end{abstract}

\maketitle

\section{Introduction}

During the last few years the van der Waals and Casimir interactions have
attracted widespread interest due to important role they play in many
physical phenomena \cite{1,2}. In most cases, however, the emphasis has
been made on the forces acting between two closely spaced bodies, be it
two atoms or molecules, an atom or a molecule and a macroscopic surface,
or two macroscopic surfaces. It is common knowledge that the van der Waals
and Casimir forces are caused by the zero-point and thermal fluctuations
of the electromagnetic field and are described by the Lifshitz theory
of dispersion forces \cite{3}. At the moment these forces are actively
investigated not only theoretically, but also experimentally (see Refs.
\cite{4,5} for a review) and are used in technological applications
\cite{6,7,8}.

Another important role of dispersion interactions is that they contribute
to the free energy of free-standing material films and films deposited on
some material plates. The formulation of this problem goes back to
Derjaguin who took into account the dispersion-force contribution in
studies of stability of thin films and introduced the concept of
disjoining pressure (see Refs. \cite{9,10} for a review). During a few
decades this contribution to the free energy, which depends on the film
thickness, was estimated using the power-type force law and the Hamaker
constant.

In the present state of the art, the question of the Casimir energy for
a free-standing or sandwiched between two dielectric plates metallic
film was raised in Ref. \cite{11}. Then, the Casimir energy of a
free-standing in vacuum metallic film was considered in Refs. \cite{11a,11b}.
In doing so, the dielectric properties of metal were described by either
the Drude or the plasma model. When employing the plasma model, the
Lifshitz theory at nonzero temperature has been used in calculations.
However, all calculations employing the Drude model have been performed
at zero temperature. This did not allow to reveal significant differences
in theoretical results for the free energy of metallic films predicted by
the Lifshitz theory combined with either the Drude or the plasma model.

Full investigation of the Casimir free energy and pressure for metallic
films in the framework of the Lifshitz theory at nonzero temperature
was performed in Refs. \cite{12,13,14}. The cases of a free-standing
or sandwiched between two dielectric plates \cite{12}, deposited on a
metal plate \cite{13} or made of magnetic metal \cite{14} metallic films
have been considered. The dielectric properties of metals were described
by using the optical data for the complex index of refraction extrapolated
to zero frequency by the Drude or plasma models. It was shown that
magnitudes of the free energy of metallic films of less than 150 nm
thickness differ by up to a factor of 1000 depending on the calculation
approach used \cite{12,13,14}. So great difference is explained by the
fact that the Casimir free energy of metallic films drops to zero
exponentially fast when the plasma model is used for extrapolation and
goes to the classical limit when the optical data are extrapolated by
the Drude model \cite{12,13,14}. This limit is already reached for the
film of 150 nm thickness.

Here we note that although routinely it is quite natural to use the Drude
model for extrapolation of the optical data to lower frequencies because
it takes into account the relaxation properties of conduction electrons,
there are also strong reasons for using the lossless plasma model for
this purpose in the case of fluctuating fields. The point is that the
measurement data of all precise experiments on measuring the Casimir
interaction between two material bodies separated with a vacuum gap
exclude theoretical predictions of the Lifshitz theory combined with
the Drude model and are consistent with predictions of the same theory
using the plasma model \cite{15,16,17,18,19,20,21}. For the gap width
below 1 $\mu$m, used in these experiments, the variation in theoretical
predictions of both approaches is below a few percent. Recently, however,
the differential force measurement scheme has been proposed \cite{22,23,24},
where this variation is by up to a factor 1000. The results of one of these
experiments, already performed \cite{25,26}, exclude with certainty the
predictions of the Drude model and are consistent with the plasma model.
Basing on this, it was hypothesized that reaction of a physical system
to real and fluctuating electromagnetic fields (having a nonzero and zero
expectation values, respectively) might be different \cite{14,27}.

On theoretical side, it was shown \cite{28,29} that for two metallic
plates, separated by more than 6 $\mu$m distance, the classical statistical
physics predicts the same Casimir force as does the Lifshitz theory
combined with the plasma model. By contrast, for metals with perfect
crystal lattices the Lifshitz theory was shown to violate the third low
of thermodynamics (the Nernst heat theorem) when the Drude model is used
\cite{30,31,32,33,34}. In this respect, one may guess that even at
separations exceeding 6 $\mu$m, where the major contribution to the
Casimir force between two parallel plates becomes classical, the quantum
effects still remain important and make the classical treatment
inapplicable.

In view of the above problem, which is often called ``the Casimir puzzle",
it is desirable to present additional arguments regarding an
applicability of the Drude and plasma models in calculations of the
Casimir free energy of metallic films. Here, the calculation results differ
greatly, and the subject is not of only an academic character because
the obtained values should be taken into account in the conditions of
film stability.

In this paper, we derive the asymptotic expressions  at low temperature
for thermal
corrections to the Casimir free energy and pressure of metallic films
described by the plasma model. The asymptotic
behavior of the Casimir entropy is also obtained. Unlike the familiar
case of two parallel plates separated with a gap, all these quantities
decrease exponentially fast with increasing film thickness and do not
have the classical limit by depending on $\hbar$ at arbitrarily large
film thicknesses. It is shown that the Casimir entropy of a film
preserves the positive values and, in the limiting case of zero
temperature, goes to zero. Thus, it is proved that the Casimir entropy
of metallic films described by the plasma model satisfies the Nernst
heat theorem, i.e., the Lifshitz theory is thermodynamically consistent.

Then, the low-temperature behavior of the Casimir free energy and entropy
for metallic films described by the Drude model is considered. We show
that in the limiting case of zero temperature the Casimir entropy goes
to a positive value depending on the parameters of a film. Therefore,
the Nernst heat theorem is violated \cite{35,36}. Furthermore, it is
demonstrated that in this case the Casimir free energy does not go to
zero in the limiting case of ideal metal film, which is in
contradiction to the fact that electromagnetic oscillations cannot
penetrate in an interior of ideal metal. Thus, the description of a
film metal by the Drude model in the Lifshitz theory results in
violation of basic thermodynamic demands. Because of this, the
dispersion-force contribution to the free energy of metallic films
might need a reconsideration taking into account that the low-frequency
behavior of the film metal is described by the plasma model.

The paper is organized as follows. In Sec. II, we present general
formalism and derive the low-temperature behavior of the Casimir free
energy, pressure and entropy for metallic films described by the
plasma model. In Sec. III, we consider the low-temperature behavior of the
Casimir free energy and entropy of metallic films with perfect crystal
lattices described by the Drude model and demonstrate violation of the
Nernst heat theorem. Section IV contains our conclusions and discussion.
In Appendix, some details of the mathematical derivations are presented.

\section{Metals described by the plasma model}

The free energy per unit area of a free-standing metallic film of
thickness $a$ in vacuum at temperature $T$ in thermal equilibrium with
an environment is given by the Lifshitz formula \cite{2,3}
\begin{eqnarray}
&&
{\cal F}(a,T)=\frac{k_BT}{2\pi}\sum_{l=0}^{\infty}{\vphantom{\sum}}^{\prime}
\int_{0}^{\infty}k_{\bot}dk_{\bot}
\label{eq1}\\
&&~~~~~~~~~
\times\sum_{\alpha}\ln\left[1-
r_{\alpha}^2(i\xi_l,k_{\bot})e^{-2ak(i\xi_l,k_{\bot})}\right].
\nonumber
\end{eqnarray}
\noindent
Here, $k_B$ is the Boltzmann constant, $k_{\bot}$ is the magnitude of the
projection of the wave vector on the film plane, $\xi_l=2\pi k_BTl/\hbar$,
$l=0,\,1,\,2,\,\ldots$ are the Matsubara frequencies, the prime on the
summation sign multiplies the term with $l=0$ by 1/2, and
\begin{equation}
k(i\xi_l,k_{\bot})=\sqrt{k_{\bot}^2+\varepsilon_l\frac{\xi_l^2}{c^2}},
\label{eq2}
\end{equation}
\noindent
where $\varepsilon_l\equiv\varepsilon(i\xi_l)$ is the frequency-dependent
dielectric permittivity of film metal calculated at the pure imaginary
Matsubara frequencies.

The reflection coefficients for two independent polarizations of the
electromagnetic field, transverse magnetic ($\alpha={\rm TM}$) and transverse
electric ($\alpha={\rm TE}$), are given by
\begin{eqnarray}
&&
r_{\rm TM}(i\xi_l,k_{\bot})=\frac{k(i\xi_l,k_{\bot})-\varepsilon_l
q(i\xi_l,k_{\bot})}{k(i\xi_l,k_{\bot})+\varepsilon_l
q(i\xi_l,k_{\bot})},
\nonumber \\
&&
r_{\rm TE}(i\xi_l,k_{\bot})=\frac{k(i\xi_l,k_{\bot})-
q(i\xi_l,k_{\bot})}{k(i\xi_l,k_{\bot})+q(i\xi_l,k_{\bot})},
\label{eq3}
\end{eqnarray}
\noindent
where
\begin{equation}
q(i\xi_l,k_{\bot})=\sqrt{k_{\bot}^2+\frac{\xi_l^2}{c^2}}.
\label{eq4}
\end{equation}

Equation (\ref{eq1}) is obtained \cite{12} from the standard Lifshitz formula for
a three-layer system \cite{37,38,39}, where the metallic plate is sandwiched between
two vacuum semispaces. Note that the reflection coefficients (\ref{eq3}) have the
opposite sign, as compared to the case of two plates separated by the vacuum gap
\cite{2}. The reason is that here an incident wave inside the film material goes
to its boundary plane with a vacuum, and not from the vacuum gap to the material
boundary. Another distinctive feature of Eq.~(\ref{eq1}) from the standard Lifshitz
formula is that here the dielectric permittivity of metal enters the power of the
exponent [in the standard case this exponent contains the quantity $q$ defined in
Eq.~(\ref{eq4})]. This makes the properties of the free energy (\ref{eq1}) quite different
from those in the case of two parallel plates separated by a vacuum gap.

It is convenient to introduce the dimensionless integration variable
\begin{equation}
y=2aq(i\xi_l,k_{\bot}).
\label{eq5}
\end{equation}
\noindent
Using the characteristic frequency $\omega_c\equiv c/(2a)$, we also pass on the
dimensionless Matsubara frequencies
\begin{equation}
\zeta_l=\frac{\xi_l}{\omega_c}=4\pi\frac{k_BTa}{\hbar c}l\equiv\tau l.
\label{eq6}
\end{equation}
\noindent
Then, the Casimir free energy (\ref{eq1}) takes the form
\begin{eqnarray}
&&
{\cal F}(a,T)=\frac{k_BT}{8\pi a^2}\sum_{l=0}^{\infty}{\vphantom{\sum}}^{\prime}
\int_{\zeta_l}^{\infty}y\,dy
\label{eq7}\\
&&~~~~~~~~~
\times\sum_{\alpha}\ln\left[1-
r_{\alpha}^2(i\zeta_l,y)e^{-\sqrt{y^2+(\varepsilon_l-1)\zeta_l^2}}\right].
\nonumber
\end{eqnarray}
\noindent
In terms of the quantities (\ref{eq5}) and (\ref{eq6}), the reflection coefficients
(\ref{eq3}) are given by
\begin{eqnarray}
&&
r_{\rm TM}(i\zeta_l,y)=\frac{\sqrt{y^2+(\varepsilon_l-1)\zeta_l^2}-\varepsilon_l
y}{\sqrt{y^2+(\varepsilon_l-1)\zeta_l^2}+\varepsilon_l y},
\nonumber \\
&&
r_{\rm TE}(i\zeta_l,y)=\frac{\sqrt{y^2+(\varepsilon_l-1)\zeta_l^2}-
y}{\sqrt{y^2+(\varepsilon_l-1)\zeta_l^2}+ y}.
\label{eq8}
\end{eqnarray}

Now we assume that at the imaginary Matsubara frequencies the film metal is described
by the lossless plasma model
\begin{equation}
\varepsilon_{l,p}=1+\frac{\omega_p^2}{\xi_l^2},
\label{eq9}
\end{equation}
\noindent
where $\omega_p$ is the plasma frequency. In terms of dimensionless frequencies (\ref{eq6}),
the dielectric permittivity (\ref{eq9}) takes the form
\begin{equation}
\varepsilon_{l,p}=1+\frac{\tilde{\omega}_p^2}{\zeta_l^2},
\quad
\tilde{\omega}_p\equiv\frac{\omega_p}{\omega_c}=\frac{2a\omega_p}{c}.
\label{eq10}
\end{equation}

Substituting Eq.~(\ref{eq10}) in  Eq.~(\ref{eq8}), one obtains the reflection coefficients
in the case when the plasma model is used
\begin{eqnarray}
&&
r_{{\rm TM},p}(i\zeta_l,y)=\frac{\zeta_l^2(\sqrt{y^2+\tilde{\omega}_p^2}-y)-\tilde{\omega}_p^2
y}{\zeta_l^2(\sqrt{y^2+\tilde{\omega}_p^2}+y)+\tilde{\omega}_p^2y},
\nonumber \\
&&
r_{{\rm TE},p}(i\zeta_l,y)=r_{{\rm TE},p}(y)=
\frac{\sqrt{y^2+\tilde{\omega}_p^2}-y}{\sqrt{y^2+\tilde{\omega}_p^2}+y}.
\label{eq11}
\end{eqnarray}

For the film described by the plasma model, it is convenient to rewrite the Casimir free
energy (\ref{eq7}) as
\begin{equation}
{\cal F}_p(a,T)=\frac{k_BT}{8\pi a^2}\sum_{l=0}^{\infty}{\vphantom{\sum}}^{\prime}
\Phi(\zeta_l)=\frac{k_BT}{8\pi a^2}\sum_{l=0}^{\infty}{\vphantom{\sum}}^{\prime}
[\Phi_{\rm TM}(\zeta_l)+\Phi_{\rm TE}(\zeta_l)],
\label{eq12}
\end{equation}
\noindent
where
\begin{equation}
\Phi_{\rm TM(TE)}(x)=
\int_{x}^{\infty}\!\!\!\!y\,dy
\ln\left[1-
r_{{\rm TM(TE)},p}^2(ix,y)e^{-\sqrt{y^2+\tilde{\omega}_p^2}}\right].
\label{eq13}
\end{equation}

It is well known that the Casimir free energy can be presented in the form
\begin{equation}
{\cal F}_p(a,T)=E_p(a,T)+\Delta_T{\cal F}_p(a,T),
\label{eq14}
\end{equation}
\noindent
where the Casimir energy per unit area at zero temperature is given by \cite{2,3}
\begin{equation}
E_p(a,T)=\frac{\hbar c}{32\pi^2a^3}\int_{0}^{\infty}\!\!d\zeta\Phi(\zeta)
\label{eq15}
\end{equation}
\noindent
and $\Delta_T{\cal F}_p$ is the thermal correction to it.

Applying the Abel-Plana formula to Eq.~(\ref{eq12}) and taking into account that
$\zeta_l=\tau l$, one arrives at
\begin{equation}
\Delta_T{\cal F}_p(a,T)=i\frac{k_BT}{8\pi a^2}\int_{0}^{\infty}\!\!dt
\frac{\Phi(i\tau t)-\Phi(-i\tau t)}{e^{2\pi t}-1}.
\label{eq16}
\end{equation}
\noindent
It is evident that the low-temperature behavior of the Casimir free energy of thin
metallic films can be found from the perturbation expansion of Eq.~(\ref{eq16})
under the condition $\tau t\ll 1$. In doing so, it is convenient to consider the
contributions of the TM and TE modes to Eq.~(\ref{eq16}) separately taking into account
Eq.~(\ref{eq12}). Note that for two media with a gap in-between the low 
temperature expansion in the Lifshitz formula was performed 
in Refs.{\ }\cite{30,31,32,33,34}. These results were systemized and partly 
extended in Ref.{\  }\cite{34a}.

We start from the TE mode because in this case the function under the integral in
Eq.~(\ref{eq13}) does not depend on $x$ due to the second equality in Eq.~(\ref{eq11}).
This mean that the total dependence of $\Phi_{\rm TE}(x)$ on $x$ is determined by only
the lower integration limit in Eq.~(\ref{eq13}).

Now we expand the function  $\Phi_{\rm TE}(x)$ in
a series in powers of $x$. The first term in this series is
\begin{equation}
\Phi_{\rm TE}(0)=\int_{0}^{\infty}\!\!\!ydy\ln[1-r_{{\rm TE},p}^2(y)
e^{-\sqrt{y^2+\tilde{\omega}_p^2}}].
\label{eq17}
\end{equation}
\noindent
This is a converging integral, which does not contribute to the difference
\begin{equation}
\Delta\Phi_{\rm TE}\equiv\Phi_{\rm TE}(i\tau t)-\Phi_{\rm TE}(-i\tau t),
\label{eq18}
\end{equation}
\noindent
entering Eq.~(\ref{eq16}).

Then, calculating the first and second derivatives of Eq.~(\ref{eq13}), one finds
\begin{equation}
\Phi_{\rm TE}^{\prime}(0)=0,\qquad
\Phi_{\rm TE}^{\prime\prime}(0)=-\ln(1-e^{-\tilde{\omega}_p}).
\label{eq19}
\end{equation}
\noindent
The respective terms of the power series again do not contribute to the difference
(\ref{eq18}).

Finally, we find
\begin{equation}
\Phi_{\rm TE}^{\prime\prime\prime}(0)=-\frac{8}{\tilde{\omega}_p}\,
\frac{1}{e^{\tilde{\omega}_p}-1}
\label{eq20}
\end{equation}
\noindent
and, thus,
\begin{equation}
\Phi_{\rm TE}(x)=\Phi_{\rm TE}(0)-\frac{x^2}{2}\ln(1-e^{-\tilde{\omega}_p})
-\frac{4}{3\tilde{\omega}_p}\,
\frac{x^3}{e^{\tilde{\omega}_p}-1}+O(x^4),
\label{eq21}
\end{equation}
\noindent
where $\Phi_{\rm TE}(0)$ is defined in Eq.~(\ref{eq17}).

Restricting ourselves by the third perturbation order, Eqs.~(\ref{eq18}) and
(\ref{eq21}) result in
\begin{equation}
\Delta\Phi_{\rm TE}\approx i\frac{8}{3\tilde{\omega}_p}\,
\frac{\tau^3t^3}{e^{\tilde{\omega}_p}-1}.
\label{eq22}
\end{equation}

We are coming now to the contribution of the TM mode to the quantity (\ref{eq16}).
This case is more complicated because both the lower integration limit and the
function under the integral in Eq.~(\ref{eq13}) depend on $x$.

By calculating several first derivatives of Eq.~(\ref{eq13}), where the reflection coefficient
is defined by the first equality in Eq.~(\ref{eq11}), one finds
\begin{eqnarray}
&&
\Phi_{\rm TM}(0)=\int_{0}^{\infty}\!\!\!ydy\ln(1-
e^{-\sqrt{y^2+\tilde{\omega}_p^2}}),
\nonumber \\
&&
\Phi_{\rm TM}^{\prime}(0)=0,
\label{eq23}\\
&&
\Phi_{\rm TM}^{\prime\prime}(0)=\frac{8}{\tilde{\omega}_p^2}
\int_{0}^{\infty}\!\!\!dy
\frac{\sqrt{y^2+\tilde{\omega}_p^2}}{e^{\sqrt{y^2+\tilde{\omega}_p^2}}-1}
-\ln(1-e^{-\tilde{\omega}_p}),
\nonumber \\
&&
\Phi_{\rm TM}^{\prime\prime\prime}(0)=-\frac{16}{\tilde{\omega}_p}\,
\frac{1}{e^{\tilde{\omega}_p}-1}.
\nonumber
\end{eqnarray}

It is evident that the first two terms in the power series, defined by Eq.~(\ref{eq23}),
\begin{equation}
\Phi_{\rm TM}(x)=\Phi_{\rm TM}(0)+\frac{\Phi_{\rm TM}^{\prime\prime}(0)x^2}{2}
-\frac{8}{3\tilde{\omega}_p}\,
\frac{x^3}{e^{\tilde{\omega}_p}-1}+O(x^4),
\label{eq24}
\end{equation}
\noindent
do not contribute to the quantity
\begin{equation}
\Delta\Phi_{\rm TM}\equiv\Phi_{\rm TM}(i\tau t)-\Phi_{\rm TM}(-i\tau t).
\label{eq25}
\end{equation}

Then, restricting ourselves by the third perturbation order, we arrive at
\begin{equation}
\Delta\Phi_{\rm TM}\approx i\frac{16}{3\tilde{\omega}_p}\,
\frac{\tau^3t^3}{e^{\tilde{\omega}_p}-1}.
\label{eq26}
\end{equation}

By summing up Eqs.~(\ref{eq22}) and (\ref{eq26}), one obtains
\begin{equation}
\Phi(i\tau t)-\Phi(-i\tau t)=
\Delta\Phi_{\rm TM}+\Delta\Phi_{\rm TM}\approx i\frac{8}{\tilde{\omega}_p}\,
\frac{\tau^3t^3}{e^{\tilde{\omega}_p}-1}.
\label{eq27}
\end{equation}

Substituting this result in Eq.~(\ref{eq16}), integrating with respect to $t$ and
returning to the dimensional variables, we find the behavior of the thermal
correction to the Casimir energy of metallic film at low temperature
\begin{equation}
\Delta_T{\cal F}_p(a,T)=-\frac{2\pi^2(k_BT)^4}{15\hbar^3c^2\omega_p
(e^{2a\omega_p/c}-1)}.
\label{eq28}
\end{equation}

The respective thermal correction to the Casimir pressure of a free-standing
metallic film at low $T$ takes the form
\begin{equation}
\Delta_TP_p(a,T)=-\frac{\partial{\cal F}_p(a,T)}{\partial a}
=-\frac{4\pi^2(k_BT)^4}{15\hbar^3c^3}
\frac{e^{2a\omega_p/c}}{(e^{2a\omega_p/c}-1)^2}.
\label{eq29}
\end{equation}

An interesting feature of Eqs.~(\ref{eq28}) and (\ref{eq29}) is that the thermal
corrections to the Casimir energy and pressure of metallic film, calculated using
the plasma model, go to zero exponentially fast with increasing film thickness $a$.
Thus, there is no classical limit in this case.

Another important point is that for fixed film thickness the Casimir free energy
and pressure of the film go to zero in the limiting case $\omega_p\to\infty$.
This is true for both the thermal corrections (\ref{eq28}) and (\ref{eq29})
and for the zero-temperature quantities $E(a)$ and $P(a)$. Note that
for $\omega_p\to\infty$ the magnitudes
of both the TM and TE reflection coefficients (\ref{eq11}) go to unity, i.e., the film
becomes perfectly reflecting. One can conclude that when the plasma model is used
in calculations an ideal metal film is characterized by the zero Casimir energy and
pressure, as it should be because the electromagnetic fluctuations cannot penetrate
in an interior of ideal metal.

{}From Eq.~(\ref{eq28}) one can also obtain the low-temperature behavior of the
Casimir entropy of metallic film
\begin{equation}
S_p(a,T)=- \frac{\partial{\cal F}_p(a,T)}{\partial T}=
\frac{8\pi^2k_B(k_BT)^3}{15\hbar^3c^2\omega_p
(e^{2a\omega_p/c}-1)}.
\label{eq30}
\end{equation}
\noindent
It is seen that the Casimir entropy of a film is positive. When the temperature
vanishes, one has from Eq.~(\ref{eq30})
\begin{equation}
S_p(a,T)\to 0,
\label{eq31}
\end{equation}
\noindent
i.e., the Casimir entropy of metallic film calculated using the plasma model
satisfies the Nernst heat theorem.

In the end of this section, we discuss the application region of asymptotic
Eqs.~(\ref{eq28})--(\ref{eq30}), which were derived under a condition
$x\ll 1$, i.e., $\tau t\ll 1$. Taking into account that the dominant contribution
to the integral (\ref{eq16}) is given by $t\sim 1/(2\pi)$ and considering the
definition of $\tau$ in Eq.~(\ref{eq6}), one rewrites the application condition
in the form
\begin{equation}
 k_BT\ll\frac{\hbar c}{2a}=\hbar\omega_c.
\label{eq32}
\end{equation}
\noindent
For a typical film thickness $a=100\,$nm, this inequality results in
$T\ll 11400\,$K, i.e., Eqs.~(\ref{eq28})--(\ref{eq30}) are well applicable under
a condition $T\leqslant 1000\,$K. With increasing film thickness the application
region of Eqs.~(\ref{eq28})--(\ref{eq30}) becomes more narrow, For example,
for $a=1\,\mu$m these equations are applicable at $T\leqslant 100\,$K.

\section{Metals described by the Drude model}

Now we describe metal of the film by the Drude model which takes into account the
relaxation properties of conduction electrons. At the pure imaginary Matsubara
frequencies the dielectric permittivity of the Drude metal takes the form
\begin{equation}
\varepsilon_{l,D}=1+\frac{\omega_p^2}{\xi_l[\xi_l+\gamma(T)]},
\label{eq33}
\end{equation}
\noindent
where $\gamma(T)$ is the temperature-dependent relaxation parameter.

Using the dimensionless variables (\ref{eq6}) and (\ref{eq10}) and introducing
the dimensionless relaxation parameter,
\begin{equation}
\tilde{\gamma}(T)=\frac{\gamma(T)}{\omega_c},
\label{eq34}
\end{equation}
\noindent
Eq.~(\ref{eq33}) can be rewritten as
\begin{equation}
\varepsilon_{l,D}=1+\frac{\tilde{\omega}_p^2}{\zeta_l[\zeta_l+\tilde{\gamma}(T)]}.
\label{eq35}
\end{equation}

It is convenient also to introduce one more dimensionless parameter
\begin{equation}
\delta_l(T)=\frac{\tilde{\gamma}(T)}{\zeta_l}=\frac{\gamma(T)}{\xi_l}=
\frac{\hbar\gamma(T)}{2\pi k_BT}\,\frac{1}{l},
\label{eq36}
\end{equation}
\noindent
where $l\geqslant 1$.

It is easily seen that for metals with perfect crystal lattices this parameter
satisfies a condition
\begin{equation}
\delta_l(T)\ll 1,
\label{eq37}
\end{equation}
\noindent
and becomes progressively smaller with decreasing temperature.
Thus, at $T=300\,$K for good metals we have $\gamma\sim 10^{13}\,$rad/s
(for Au $\gamma=5.3\times 10^{13}\,$rad/s), whereas
$\xi_1=2.5\times 10^{14}\,$rad/s. In the temperature region
$T_D/4<T<300\,$K, where $T_D$ is the Debye temperature (for Au we have
$T_D=165\,$K \cite{40}), it holds $\gamma(T)\sim T$, i.e., the value of
$\delta_l$ remains unchanged.
In the region from $T_D/4$ down to liquid  helium temperature $\gamma(T)\sim T^5$
in accordance to the Bloch-Gr\"{u}neisen law \cite{41} and at lower temperatures
$\gamma(T)\sim T^2$ for metals with perfect crystal lattices \cite{40}.
As a result, even the quantity $\delta_1(T)$ and, all the more, $\delta_l(T)$
go to zero when $T$ vanishes. For example, for Au at $T=30$ and 10\,K one has
$\delta_1\approx 5\times 10^{-2}$ and $2\times 10^{-3}$, respectively.

Now we express the permittivity (\ref{eq35}) in terms of the small parameter (\ref{eq37})
\begin{equation}
\varepsilon_{l,D}=1+\frac{\tilde{\omega}_p^2}{\zeta_l^2[1+\delta_l(T)]}
\label{eq38}
\end{equation}
\noindent
and, in the first perturbation order in this parameter, obtain
\begin{equation}
\varepsilon_{l,D}\approx\varepsilon_{l,p}-
\frac{\tilde{\omega}_p^2}{\zeta_l^2}\,\delta_l(T).
\label{eq39}
\end{equation}

We next use the following identical representation for the Casimir free energy of
metallic film calculated using the Drude model:
\begin{equation}
{\cal F}_D(a,T)={\cal F}_p(a,T)+{\cal F}_D^{(0)}(a,T)-{\cal F}_p^{(0)}(a,T)
+{\cal F}^{(\gamma)}(a,T).
\label{eq40}
\end{equation}
\noindent
Here, ${\cal F}_p$ is the free energy (\ref{eq12}) calculated using the plasma
model and ${\cal F}_p^{(0)}$ is its zero-frequency term
\begin{eqnarray}
&&
{\cal F}_p^{(0)}(a,T)=\frac{k_BT}{16\pi a^2}\int_{0}^{\infty}\!\!\!ydy
\left\{\vphantom{\ln\left[r_{{\rm TE},p}^2e^{-\sqrt{y^2+\tilde{\omega}_p^2}}\right]}
\ln\left(1-e^{-\sqrt{y^2+\tilde{\omega}_p^2}}\right)\right.
\nonumber \\
&&~~~~
\left.+\ln\left[1-r_{{\rm TE},p}^2(y)
e^{-\sqrt{y^2+\tilde{\omega}_p^2}}\right]\right\},
\label{eq41}
\end{eqnarray}
\noindent
where the reflection coefficient $r_{{\rm TE},p}$ is defined in the second line of
Eq.~(\ref{eq11}).

The quantity ${\cal F}_D^{(0)}$ in Eq.~(\ref{eq40}) is the zero-frequency term
in the Casimir free energy of a film when the Drude model is used in calculations.
{}From Eqs.~(\ref{eq7}) and (\ref{eq8}) one obtains
\begin{eqnarray}
&&
{\cal F}_D^{(0)}(a,T)=\frac{k_BT}{16\pi a^2}\int_{0}^{\infty}\!\!\!ydy
\ln\left(1-e^{-y}\right)
\nonumber \\
&&~~~~~~~~
=-\frac{k_BT}{16\pi a^2}\,\zeta(3),
\label{eq42}
\end{eqnarray}
\noindent
where $\zeta(z)$ is the Riemann zeta function.

Finally, the quantity  ${\cal F}^{(\gamma)}$ in Eq.~(\ref{eq40}) is the difference
of all nonzero-frequency Matsubara terms in the Casimir free energy (\ref{eq7})
calculated using the Drude and plasma models
\begin{eqnarray}
&&
{\cal F}^{(\gamma)}(a,T)=\frac{k_BT}{8\pi a^2}\sum_{l=1}^{\infty}
\int_{\zeta_l}^{\infty}\!\!\!ydy
\label{eq43}  \\
&&~~\times
\left\{\ln\left[1-r_{{\rm TM},D}^2(i\zeta_l,y)e^{-\sqrt{y^2+\tilde{\omega}_p^2(1-\delta_l)}}\right]
\right.
\nonumber\\
&&~~~~~~
+\ln\left[1-r_{{\rm TE},D}^2(i\zeta_l,y)e^{-\sqrt{y^2+\tilde{\omega}_p^2(1-\delta_l)}}\right]
\nonumber \\
&&~~~~~~
-\ln\left[1-r_{{\rm TM},p}^2(i\zeta_l,y)e^{-\sqrt{y^2+\tilde{\omega}_p^2}}\right]
\nonumber \\
&&~~~~~~
\left.-\ln\left[1-r_{{\rm TE},p}^2(i\zeta_l,y)
e^{-\sqrt{y^2+\tilde{\omega}_p^2}}\right]\right\},
\nonumber
\end{eqnarray}

As shown in Appendix,
\begin{equation}
\lim_{T\to 0}{\cal F}^{(\gamma)}(a,T)=0,\quad
\lim_{T\to 0}\frac{\partial{\cal F}^{(\gamma)}(a,T)}{\partial T}=0.
\label{eq44}
\end{equation}
\noindent
Because of this, we concentrate our attention on the other contributions to the
right-hand side of Eq.~(\ref{eq40}).

The quantity ${\cal F}_p$ is already found in Eqs.~(\ref{eq14}) and (\ref{eq28}),
and the quantity ${\cal F}_D^{(0)}$ is presented in Eq.~(\ref{eq42}). Here, we
calculate the quantity ${\cal F}_p^{(0)}$ defined in Eq.~(\ref{eq41}).
Let us start with the integral
\begin{equation}
I_1(\tilde{\omega}_p)\equiv\int_{0}^{\infty}\!\!\!y\,dy
\ln\left(1-e^{-\sqrt{y^2+\tilde{\omega}_p^2}}\right).
\label{eq45}
\end{equation}

Expanding the logarithm in power series and introducing the new integration variable
\begin{equation}
t=n\sqrt{y^2+\tilde{\omega}_p^2},
\label{eq46}
\end{equation}
\noindent
one obtains from Eq.~(\ref{eq45})
\begin{eqnarray}
&&
I_1(\tilde{\omega}_p)=-\sum_{n=1}^{\infty}\frac{1}{n^3}
\int_{n\tilde{\omega}_p}^{\infty}tdte^t
\label{eq47} \\
&&
\phantom{I_1(\tilde{\omega}_p)}
=-\sum_{n=1}^{\infty}\frac{1}{n^3}
(1+n\tilde{\omega}_p)e^{-n\tilde{\omega}_p}.
\nonumber
\end{eqnarray}
\noindent
After a summation, Eq.~(\ref{eq47}) results in
\begin{equation}
I_1(\tilde{\omega}_p)=-\left[{\rm Li}_3(e^{-\tilde{\omega}_p})+
\tilde{\omega}_p{\rm Li}_2(e^{-\tilde{\omega}_p})\right],
\label{eq48}
\end{equation}
where ${\rm Li}_k(z)$ is the polylogarithm function.

Now we consider the second integral entering Eq.~(\ref{eq41}), i.e.,
\begin{equation}
I_2(\tilde{\omega}_p)\equiv\int_{0}^{\infty}\!\!\!y\,dy
\ln\left[1-r_{{\rm TE},p}^2(y)e^{-\sqrt{y^2+\tilde{\omega}_p^2}}\right],
\label{eq49}
\end{equation}
\noindent
where the reflection coefficient $r_{{\rm TE},p}$ is defined in Eq.~(\ref{eq11}).
Note that for physical values of $\tilde{\omega}_p$ the quantity subtracted from
unity under the logarithm in Eq.~(\ref{eq49}) is much smaller than unity. The reason is that
if $\tilde{\omega}_p$ is not large the squared reflection coefficient $r_{{\rm TE},p}^2$
is rather small. Then, one can expand the logarithm up to the first power of this
parameter and obtain
\begin{equation}
I_2(\tilde{\omega}_p)\approx-\int_{0}^{\infty}\!\!\!y\,dy\,
r_{{\rm TE},p}^2(y)e^{-\sqrt{y^2+\tilde{\omega}_p^2}}.
\label{eq50}
\end{equation}

Numerical computations show that Eqs.~(\ref{eq49}) and (\ref{eq50}) lead to nearly
coincident results for $\tilde{\omega}_p\geqslant 0.5$. Taking into account the
definition of $\tilde{\omega}_p$ in Eq.~(\ref{eq10}), this results in the condition
$a\geqslant 5.4\,$nm for a thickness of Au film with $\omega_p=1.37\times 10^{16}\,$rad/s.
This is quite sufficient for our purposes because here we consider metallic films of
more than 7\,nm thickness, which can be described by the isotropic dielectric
permittivity \cite{42} (for thinner Au films the effect of anisotropy should be taken
into account \cite{43}).

Now we introduce the variable $t=y/\tilde{\omega}_p$ and, using Eq.~(\ref{eq11}),
identically represent the quantity $r_{{\rm TE},p}^2$ in the form
\begin{equation}
r_{{\rm TE},p}^2(y)=1+8t^2+8t^4-4t\sqrt{1+t^2}-8t^2\sqrt{1+t^2}.
\label{51}
\end{equation}
\noindent
Introducing the integration variable $t$ in Eq.~(\ref{eq50}), one finds
\begin{eqnarray}
&&
I_2(\tilde{\omega}_p)\approx -\tilde{\omega}_p^2\int_{0}^{\infty}\!\!\!t\,dt
e^{-\tilde{\omega}_p\sqrt{1+t^2}}
\label{eq52}\\
&&~~~
\times(1+8t^2+8t^4-4t\sqrt{1+t^2}-8t^2\sqrt{1+t^2}).
\nonumber
\end{eqnarray}

Calculating all the five integrals in Eq.~(\ref{eq52}) \cite{44}, we arrive at
\begin{eqnarray}
&&
I_2(\tilde{\omega}_p)\approx -\left(\tilde{\omega}_p+17+\frac{112}{\tilde{\omega}_p}
+\frac{432}{\tilde{\omega}_p^2} +\frac{960}{\tilde{\omega}_p^3}
+\frac{960}{\tilde{\omega}_p^4}\right) e^{-\tilde{\omega}_p}
\nonumber \\
&&~~~
+4\left[\tilde{\omega}_pK_1(\tilde{\omega}_p)+9K_2(\tilde{\omega}_p)+
\frac{30}{\tilde{\omega}_p}K_3(\tilde{\omega}_p)\right].
\label{eq53}
\end{eqnarray}

As a result, the Casimir free energy (\ref{eq40}), calculated using the Drude model,
can be rewritten in the form
\begin{eqnarray}
&&
{\cal F}_D(a,T)={\cal F}_p(a,T)+{\cal F}^{(\gamma)}(a,T)
\label{eq54}\\
&&~~
-\frac{k_BT}{16\pi a^2}\left[
\zeta(3)+I_1\left(\frac{2a\omega_p}{c}\right)
+I_2\left(\frac{2a\omega_p}{c}\right)\right],
\nonumber
\end{eqnarray}
\noindent
where ${\cal F}_p$ and ${\cal F}^{(\gamma)}$ are presented in Eqs.~(\ref{eq14}),
(\ref{eq28}), (\ref{eq43}), and $I_1$ and $I_2$ are found in Eqs.~(\ref{eq48}),
(\ref{eq53}).

Now we calculate the negative derivative of Eq.~(\ref{eq54}) with respect to $T$
and find the limiting value of this derivative when $T$ goes to zero using Eqs.~(\ref{eq28})
and (\ref{eq44}). The result is
\begin{equation}
S_D(a,0)=\frac{k_B}{16\pi a^2}\left[
\zeta(3)+I_1\left(\frac{2a\omega_p}{c}\right)
+I_2\left(\frac{2a\omega_p}{c}\right)\right].
\label{eq55}
\end{equation}

As is seen in Eq.~(\ref{eq55}), the Casimir entropy of metallic film
at zero temperature, calculated using the Drude model, is not equal to zero
and depends on the parameters of a film (the thickness $a$ and the plasma frequency
$\omega_p$). Thus, in this case the Nernst heat theorem is violated \cite{35,36}.

Calculations using Eqs.~(\ref{eq48}) and (\ref{eq53}) show that
\begin{equation}
S_D(a,0)>0.
\label{eq56}
\end{equation}
\noindent
Thus, for $\tilde{\omega}_p=1$ (i.e., for a Au film of approximately 11\,nm thickness)
one has $I_1=-0.79575$, $I_2=-0.02456$, which leads to the number in square brackets
in Eq.~(\ref{eq55}) $C=0.38175$. For $\tilde{\omega}_p=5$ ($a=55\,$nm) the respective
results are: $I_1=-0.04049$, $I_2=-0.006684$, and $C=1.15489$. Finally, for
$\tilde{\omega}_p=15$ ($a=165\,$nm)  $I_1=-4.894\times 10^{-6}$,
$I_2=-1.5966\times 10^{-6}$, and $C=1.20205$.
We see that with increasing film thickness the magnitudes of the 
quantities $I_1$ and $I_2$ become
negligibly small, as compared with $\zeta(3)$.

\section{Conclusions and discussion}

In the foregoing, we have considered the low-temperature behavior of the
Casimir free energy, entropy and pressure of metallic films in vacuum.
It was shown that the calculation results are quite different depending
on whether the plasma or the Drude model is used to describe the
dielectric response of a film metal. If the lossless plasma model is
used, as is suggested by the results of several precise experiments on
measuring the Casimir force, we have obtained explicit analytic
expressions for the thermal corrections to the Casimir energy and
pressure and for the Casimir entropy of a film, which are applicable
over the wide temperature region down to zero temperature. These
expressions do not have a classical limit and go to zero when the
film material becomes perfectly reflecting. The Casimir entropy is
shown to be positive and satisfying the Nernst heat theorem, i.e., it
goes to zero in the limiting case of zero temperature.

If the film metal is described by the Drude model taking into account
the relaxation properties of conduction electrons at low frequencies,
the calculation results are quite different, both qualitatively and
quantitatively. In accordance to what was shown in previous work
\cite{12,13,14}, the Casimir free energy and pressure reach the
classical limit for rather thin metallic films of approximately
150 nm thickness. However, in contradiction to physical intuition,
the Casimir free energy does not go to zero in the limiting case
of ideal metal film.

We have found analytically the Casimir entropy of metallic films
with perfect crystal lattices, described by the Drude model, at
zero temperature. It is demonstrated that this quantity takes a
positive value depending on the parameters of a film, i.e., the
Nernst heat theorem is violated. Thus, the case of a free-standing
film is different from the case of two nonmagnetic metal
plates described by the Drude model interacting through a vacuum
gap. In the latter case the Nernst heat theorem is also violated if
the Drude model is used in calculations, but the Casimir entropy
takes a negative value at $T$=0  \cite{30,31,32}.

The obtained results raise a problem on what is the proper way to
calculate the dispersion-force contribution to the free energy of
metallic films. As discussed in Sec. I, the resolution of this
problem is important for investigations of stability of thin
films. Previous precise experiments on measuring the Casimir force
between metallic test bodies \cite{15,16,17,18,19,20,21,25,26}
have always been found in agreement with theoretical predictions
of the thermodynamically consistent approach using the plasma model
and excluded the theoretical predictions obtained using the Drude
model. Recently it was shown \cite{45} that theoretical description
of the Casimir interaction in graphene systems by means of the
polarization tensor, which is in agreement \cite{46} with the
experimental data \cite{47}, also satisfies the Nernst heat theorem.
Thus, there is good reason to suppose that the contribution of
dispersion forces to the free energy of metallic films should also
be calculated in a thermodynamically consistent way, i.e., using
the plasma model. An experimental confirmation to this hypothesis
might be expected within the next few years.

\section*{Acknowledgments}

The work of V.M.M. was partially supported by the Russian Government
Program of Competitive Growth of Kazan Federal University.
\appendix
\section{}

Here, we investigate the low-temperature behavior of the quantity ${\cal F}^{(\gamma)}$
defined in Eq.~(\ref{eq43}) and prove Eq.~(\ref{eq44}) used in Sec.~III.
For this purpose we expand ${\cal F}^{(\gamma)}$ up to the first order in small parameter
$\delta_l(T)$ defined in Eq.~(\ref{eq36}). According to the results of Sec.~III, for
metals with perfect crystal lattices this parameter becomes progressively smaller with
decreasing $T$.

The reflection coefficients in the case when the Drude model is used can be obtained by
substituting Eq.~(\ref{eq39}) in Eq.~(\ref{eq8})
\begin{eqnarray}
&&
r_{{\rm TM},D}(i\zeta_l,y)\approx
\frac{\sqrt{y^2+(\varepsilon_{l,p}-1)\zeta_l^2-\tilde{\omega}_p^2\delta_l}-\varepsilon_{l,p}y+
\frac{\tilde{\omega}_p^2y}{\zeta_l^2}\delta_l}{\sqrt{y^2+(\varepsilon_{l,p}-1)\zeta_l^2-
\tilde{\omega}_p^2\delta_l}+\varepsilon_{l,p}y-
\frac{\tilde{\omega}_p^2y}{\zeta_l^2}\delta_l},
\nonumber \\
&&
r_{{\rm TE},D}(i\zeta_l,y)\approx
\frac{\sqrt{y^2+(\varepsilon_{l,p}-1)\zeta_l^2-\tilde{\omega}_p^2\delta_l}-
y}{\sqrt{y^2+(\varepsilon_{l,p}-1)\zeta_l^2-\tilde{\omega}_p^2\delta_l}+y}.
\label{A1}
\end{eqnarray}

Expanding the second powers of these coefficients up to the first order of
$\delta_l=\delta_l(T)$, one obtains
\begin{eqnarray}
&&
r_{{\rm TM},D}^2(i\zeta_l,y)\approx r_{{\rm TM},p}^2(i\zeta_l,y)-
\delta_l(T)R_{\rm TM}(i\zeta_l,y),
\nonumber \\
&&
r_{{\rm TE},D}^2(i\zeta_l,y)\approx r_{{\rm TE},p}^2(i\zeta_l,y)-
\delta_l(T)R_{\rm TE}(i\zeta_l,y),
\label{A2}
\end{eqnarray}
\noindent
where the quantities $R_{\rm TM}$ and $R_{\rm TE}$ are given by
\begin{eqnarray}
&&
R_{\rm TM}(i\zeta_l,y)=\frac{2\tilde{\omega}_p^2\zeta_l^2y(\tilde{\omega}_p^2+
2y^2-\zeta_l^2)(\tilde{\omega}_p^2y+\zeta_l^2y-
\zeta_l^2\sqrt{y^2+\tilde{\omega}_p^2})}{\sqrt{y^2+\tilde{\omega}_p^2}
(\tilde{\omega}_p^2y+\zeta_l^2y+
\zeta_l^2\sqrt{y^2+\tilde{\omega}_p^2})^3},
\nonumber \\
&&
R_{\rm TE}(i\zeta_l,y)=R_{\rm TE}(y)=\frac{2\tilde{\omega}_p^2y
(\sqrt{y^2+\tilde{\omega}_p^2}-y)}{\sqrt{y^2+\tilde{\omega}_p^2}
(\sqrt{y^2+\tilde{\omega}_p^2}+y)^3}.
\label{A3}
\end{eqnarray}
\noindent
It is easily seen that for any $y\geqslant\zeta_l$ it holds
$R_{\rm TM}>0$ and $R_{\rm TE}>0$.

Now we consider the exponential factor in the first two contributions to 
Eq.~(\ref{eq43}). Up to the first order
in $\delta_l$, this factor can be presented in the form
\begin{eqnarray}
&&
e^{-\sqrt{y^2+\tilde{\omega}_p^2(1-\delta_l)}}=
e^{-\sqrt{y^2+\tilde{\omega}_p^2}
\sqrt{1-\frac{\delta_l\tilde{\omega}_p^2}{y^2+\tilde{\omega}_p^2}}}
\nonumber \\
&&~~~~~~~~~~
\approx
e^{-\sqrt{y^2+\tilde{\omega}_p^2}
\left[1-\frac{\delta_l\tilde{\omega}_p^2}{2(y^2+\tilde{\omega}_p^2)}\right]}.
\label{A4}
\end{eqnarray}

Next we use the fact that not only $\delta_l$, but also
$\delta_l\tilde{\omega}_p/2$ is the small parameters at sufficiently low
temperature. Really, in accordance to Eq.~(\ref{eq36}), the largest value of this
parameter is
\begin{equation}
\delta_1\frac{\tilde{\omega}_p}{2}=\frac{\gamma}{\xi_1}\,\frac{a\omega_p}{c}.
\label{A5}
\end{equation}
\noindent
For Au at $T=10\,$K we have $\gamma/\xi_1\approx 2\times 10^{-3}$, so that the quantity
(\ref{A5}) does not exceed 0.2 for film thicknesses $a\leqslant 2\,\mu$m.
At $T=5\,$K the parameter (\ref{A5}) does not exceed 0.2 for Au films with
 $a\leqslant 20\,\mu$m thickness.

Expanding the right-hand side of Eq.~(\ref{A4}) up to the first order in parameter
$\delta_l\tilde{\omega}_p/2$, we obtain
\begin{equation}
e^{-\sqrt{y^2+\tilde{\omega}_p^2(1-\delta_l)}}
\approx
e^{-\sqrt{y^2+\tilde{\omega}_p^2}}
\left(1+\delta_l\frac{\tilde{\omega}_p^2}{2\sqrt{y^2+\tilde{\omega}_p^2}}\right).
\label{A6}
\end{equation}

Substituting Eqs.~(\ref{A2}) and (\ref{A6}) in Eq.~(\ref{eq43}), expanding the first
two logarithms in powers  of $\delta_l$ and preserving only the terms of the first
order, one arrives at
\begin{eqnarray}
&&
{\cal F}^{(\gamma)}(a,T)\approx -\frac{k_BT}{8\pi a^2}\sum_{l=1}^{\infty}
\delta_l(t)\int_{\zeta_l}^{\infty}\!\!\!y\,dy
\nonumber \\
&&~~~
\times\left[
\frac{Q_{\rm TM}(i\zeta_l,y)}{e^{\sqrt{y^2+\tilde{\omega}_p^2}}-
r_{{\rm TM},p}^2(i\zeta_l,y)}\right.
\nonumber \\
&&~~~~~~
\left.
+\frac{Q_{\rm TE}(i\zeta_l,y)}{e^{\sqrt{y^2+\tilde{\omega}_p^2}}-
r_{{\rm TE},p}^2(i\zeta_l,y)}\right].
\label{A7}
\end{eqnarray}
\noindent
Here, we have introduced the notations
\begin{eqnarray}
&&
Q_{\rm TM}(i\zeta_l,y)=\frac{\tilde{\omega}_p^2}{2\sqrt{y^2+\tilde{\omega}_p^2}}
-R_{\rm TM}(i\zeta_l,y),
\label{A8}\\
&&
Q_{\rm TE}(i\zeta_l,y)=Q_{\rm TE}(y)=
\frac{\tilde{\omega}_p^2}{2\sqrt{y^2+\tilde{\omega}_p^2}}-R_{\rm TE}(y),
\nonumber
\end{eqnarray}
\noindent
and the quantities $R_{\rm TM}$ and $R_{\rm TE}$ are defined in Eq.~(\ref{A3}).

It is easily seen that $Q_{\rm TM}>0$ and $Q_{\rm TE}>0$, so that
${\cal F}^{(\gamma)}(a,T)<0$. This is because the magnitude of the Casimir free energy
of a film described by the Drude model is larger than that of a film described by
the plasma model (as opposed to the case of metallic plates separated with a vacuum
gap \cite{32}).

Equation (\ref{A7}) can be used to prove the validity of Eq.~(\ref{eq44}).
For this purpose we increase the magnitude of the right-hand side of Eq.~(\ref{A7})
by replacing $r_{{\rm TM(TE)},p}^2$ with unities in the denominators, and by omitting
the quantities $R_{\rm TM(TE)}$ in  Eq.~(\ref{A8}) for the numerators. Using also
the definition of $\delta_l$ in  Eq.~(\ref{eq36}), and the definition of
$\tilde{\omega}_p$ from  Eq.~(\ref{eq10}) in the prefactor, one obtains
\begin{equation}
|{\cal F}^{(\gamma)}(a,T)|<\frac{\hbar\gamma(T)\omega_p^2}{4\pi^2c^2}
\sum_{l=1}^{\infty}\frac{1}{l}\int_{\zeta_l}^{\infty}
\frac{y\,dy}{\sqrt{y^2+\tilde{\omega}_p^2}}\,
\frac{1} {e^{\sqrt{y^2+\tilde{\omega}_p^2}}-1}.
\label{A9}
\end{equation}

Now we introduce the new variable $t=\sqrt{y^2+\tilde{\omega}_p^2}$ and expanding in
powers of $e^{-t}$ find
\begin{equation}
|{\cal F}^{(\gamma)}(a,T)|<\frac{\hbar\gamma(T)\omega_p^2}{4\pi^2c^2}
\sum_{n=1}^{\infty}
\sum_{l=1}^{\infty}\frac{1}{l}\int_{\sqrt{\zeta_l^2+\tilde{\omega}_p^2}}^{\infty}
\!dt\,e^{-nt}.
\label{A10}
\end{equation}
\noindent
Calculating the integral and using the inequality
\begin{equation}
\frac{\zeta_l+\tilde{\omega}_p}{\sqrt{2}}<\sqrt{\zeta_l^2+\tilde{\omega}_p^2},
\label{A11}
\end{equation}
\noindent
we arrive at
\begin{equation}
|{\cal F}^{(\gamma)}(a,T)|<\frac{\hbar\gamma(T)\omega_p^2}{4\pi^2c^2}
\sum_{n=1}^{\infty}\frac{1}{n}
\sum_{l=1}^{\infty}\frac{1}{l}e^{-n\frac{\tilde{\omega}_p+\zeta_l}{\sqrt{2}}}.
\label{A12}
\end{equation}

Taking into account that $\zeta_l=\tau l$, we perform a summation with respect to $l$
and obtain
\begin{equation}
|{\cal F}^{(\gamma)}(a,T)|<-\frac{\hbar\gamma(T)\omega_p^2}{4\pi^2c^2}
\sum_{n=1}^{\infty}\frac{1}{n}
e^{-n\frac{\tilde{\omega}_p}{\sqrt{2}}}
\ln\left(1-e^{-n\frac{\tau}{\sqrt{2}}}\right),
\label{A13}
\end{equation}
\noindent
where, due to a smallness of $\tau$,
\begin{equation}
\ln\left(1-e^{-n\frac{\tau}{\sqrt{2}}}\right)\approx
\ln\left(n\frac{\tau}{\sqrt{2}}\right)=\ln\tau+\ln{n}-\frac{1}{2}\ln{2}.
\label{A14}
\end{equation}

Substituting Eq.~(\ref{A13}) in Eq.~(\ref{A12}), we represent the final results
in the form
\begin{equation}
|{\cal F}^{(\gamma)}(a,T)|<X(a,T),
\label{A15}
\end{equation}
\noindent
where
\begin{equation}
X(a,T)=\frac{\hbar\gamma(T)\omega_p^2}{4\pi^2c^2}\left(C_1\ln\frac{4\pi k_BTa}{\hbar c}
-C_2\right),
\label{A16}
\end{equation}
\noindent
and the following independent on $T$ coefficients are introduced
\begin{eqnarray}
&&
C_1=-\sum_{n=1}^{\infty}\frac{1}{n}
e^{-n\frac{\tilde{\omega}_p}{\sqrt{2}}}=
\ln\left(1-e^{-\frac{\tilde{\omega}_p}{\sqrt{2}}}\right),
\nonumber \\
&&
C_2=\sum_{n=1}^{\infty}\frac{2\ln{n}-\ln{2}}{2n}
e^{-n\frac{\tilde{\omega}_p}{\sqrt{2}}}.
\label{A17}
\end{eqnarray}
\noindent
Note that the second series is converging, as well as the first one.

Taking into account that for metals with perfect crystal lattices at very low
temperature $\gamma(T)\sim T^2$ (see Sec.~III), one concludes from Eq.~(\ref{A16})
that $X(a,T)\to 0$ when $T\to 0$. Then, from Eq.~(\ref{A15}), one obtains the first
equality in Eq.~(\ref{eq44}).

{}From Eq.~(\ref{A16}) it is seen that not only $X(a,0)=0$, but
\begin{equation}
\left.\frac{\partial X(a,T)}{\partial T}\right|_{T=0}=0
\label{A18}
\end{equation}
\noindent
as well. Using Eqs.~(\ref{A15}) and (\ref{A18}), one easily proves that the second
equality in Eq.~(\ref{eq44}) is valid.

In the end it is pertinent to note that the above results, including Eq.~(\ref{eq44}),
are also valid under a slower vanishing of the relaxation parameter with temperature
according to $\gamma(T)\sim T^{\beta}$ where $\beta>1$

\end{document}